\documentclass[conference,9pt]{IEEEtran}
\ifCLASSINFOpdf
\else
\fi
\usepackage{amsbsy,amscd,amsfonts,amsgen,amsmath,amsopn,amssymb,amstext,amsthm,amsxtra}
\usepackage{graphicx}\usepackage{subfigure}
\usepackage{verbatim}
\usepackage{citesort}
\usepackage{url}
\usepackage{color}
\usepackage{epstopdf}
\usepackage{caption}
\usepackage{changes}
\usepackage{bbm}
\usepackage{algorithm}
\usepackage{algpseudocode}
\usepackage[left=0.75in,right=0.75in,top=0.75in,bottom=1in]{geometry}

\graphicspath{{./} {confimg/}}

\IEEEoverridecommandlockouts

\begin{document}
\title{Approximate Message Passing \\ in Coded Aperture Snapshot Spectral Imaging}
\author{\IEEEauthorblockN{Jin Tan,$^*$ Yanting Ma,$^*$ Hoover Rueda,$^\dagger$ Dror Baron,$^*$ and Gonzalo R. Arce$^\dagger$}
\IEEEauthorblockA{$^*$Department of Electrical and Computer Engineering; NC State University; Raleigh, NC 27695, USA\\
Email: \{jtan,yma7,barondror\}@ncsu.edu\\
$^\dagger$Department of Electrical and Computer Engineering; University of Delaware;
Newark, DE 19716, USA\\Email: \{rueda,arce\}@udel.edu.}
\vspace*{0mm}
\thanks{The work of Jin Tan, Yanting Ma, and Dror Baron was supported in part by the National Science Foundation under the Grant CCF-1217749 and in part by the U.S. Army Research Office under the Contract W911NF-14-1-0314. The work of Hoover Rueda and Gonzalo R. Arce was supported by the U.S. Army Research Office under the Contract W911NF-12-1-0380.}
\vspace*{0mm}
}

%
\maketitle
\thispagestyle{empty}
\newcommand{\xhat}{\widehat{\mathbf{x}}}
\newcommand{\xhati}{\widehat{x}_i}
\def\x{{\mathbf x}}
\def\L{{\cal L}}
\begin{abstract}
We consider a compressive hyperspectral imaging reconstruction problem, where three-dimensional spatio-spectral information about a scene is sensed by a coded aperture snapshot spectral imager (CASSI). 
The approximate message passing (AMP) framework is utilized to reconstruct hyperspectral images from CASSI measurements, and 
an adaptive Wiener filter is employed as a three-dimensional image denoiser within AMP. We call our algorithm ``AMP-3D-Wiener."
The simulation results show that AMP-3D-Wiener outperforms existing widely-used algorithms such as gradient projection for sparse reconstruction (GPSR) and two-step iterative shrinkage/thresholding (TwIST) given the same amount of runtime. Moreover, in contrast to GPSR and TwIST, AMP-3D-Wiener need not tune any parameters, which simplifies the reconstruction process. 
\end{abstract}
\IEEEpeerreviewmaketitle

\vspace*{0mm}
\section{Introduction}
\label{sec:intro}
\vspace*{0mm}
{\bf Motivation:}
A hyperspectral image is a three-dimensional (3D) image cube comprised of a collection of two-dimensional (2D)
images (slices), where
each 2D image is captured at a specific wavelength. 
Hyperspectral imaging allows us to analyze spectral information about each spatial point in a scene, and has applications 
to areas such as medical imaging~\cite{Schultz2001}, remote sensing~\cite{Jia1999}, geology~\cite{Kruse2003},
and astronomy~\cite{Hege2004}. 

The imaging processes in conventional spectral imagers~\cite{Brady2009,Gat2000,Eismann2012} take a long time, because they require scanning a number of zones linearly in proportion to the desired spatial and spectral resolution.
To address the limitations of conventional spectral imaging techniques, many spectral imager sampling schemes based on compressive sensing~\cite{DonohoCS,CandesRUP,BaraniukCS2007} have been proposed~\cite{Gehm2007,Yuan2015Side,August2013Hyper}. 
The coded aperture snapshot 
spectral imager (CASSI)~\cite{Gehm2007,Wagadarikar2008,Arguello2011,Wagadarikar2008single} is a popular compressive spectral imager and acquires
image data from different wavelengths simultaneously, which significantly accelerates the imaging process. On the other hand, because
the measurements from CASSI are highly compressive, reconstructing 3D image cubes from CASSI measurements becomes challenging.
Moreover, because of the massive size of 3D image data, it is desirable to develop fast reconstruction algorithms in order to realize real time acquisition and processing.


{\bf Related work:}
Several compressive sensing algorithms have been
proposed to reconstruct image cubes from measurements acquired by CASSI. 
One of the efficient algorithms is gradient projection for sparse reconstruction (GPSR)~\cite{GPSR2007}. 
GPSR models hyperspectral image cubes as 
sparse in the Kronecker product of a 2D wavelet transform and a 1D discrete cosine transform (DCT), and
solves the $\ell_1$-minimization problem to enforce sparsity in this transform domain. 
Wagadarikar et al.~\cite{Wagadarikar2008} employed total variation~\cite{Chambolle2004} as the regularizer in the
two-step iterative shrinkage/thresholding (TwIST) framework~\cite{NewTWIST2007}, a modified and fast version of standard iterative shrinkage/thresholding. 
Apart from using the wavelet-DCT basis or total variation, one can learn a dictionary with which the image cubes can be sparsely represented~\cite{Rajwade2013,Yuan2015Side}. 
However, these algorithms all need manual tuning of some parameters, which may be time consuming.

{\bf Contributions:}
We develop a robust and fast reconstruction algorithm for CASSI using approximate message passing (AMP)~\cite{DMM2009}. AMP is an iterative algorithm that can apply image denoising at each iteration.
Previously, we proposed a 2D compressive imaging reconstruction algorithm, AMP-Wiener~\cite{Tan_CompressiveImage2014}, where an adaptive Wiener filter was applied as the image denoiser within AMP.
In this paper, 
AMP-Wiener is extended to 3D hyperspectral images, and we call it ``AMP-3D-Wiener."
Our numerical results show that
AMP-3D-Wiener reconstructs 3D image cubes with less runtime and higher quality than other reconstruction algorithms such as GPSR~\cite{GPSR2007} and TwIST~\cite{Wagadarikar2008,NewTWIST2007} (Figure~\ref{fig.iter_Psnr}), even when the regularization parameters in GPSR and TwIST have already been tuned. 
In fact, the regularization parameters in GPSR and TwIST need to be tuned carefully for each image cube, which requires to run GPSR and TwIST many times with different parameter values.
Moreover,
the improved reconstruction quality of AMP-3D-Wiener allows to reduce the number of shots taken by CASSI by a factor of $2$ (Figure~\ref{fig.shots_Psnr}).

The remainder of the paper is arranged as follows. We review 
CASSI in Section~\ref{sec:CASSI}, and describe our AMP based compressive hyperspectral imaging reconstruction algorithm in Section \ref{sec:Algo}.
Numerical results are presented in Section \ref{sec:NumSim},
while Section~\ref{sec:con} concludes.

\vspace*{0mm}
\section{Coded Aperture Snapshot Spectral Imager}
\label{sec:CASSI}
\vspace*{0mm}

\begin{figure}[t]
\vspace*{-6mm}
\includegraphics[width=85mm]{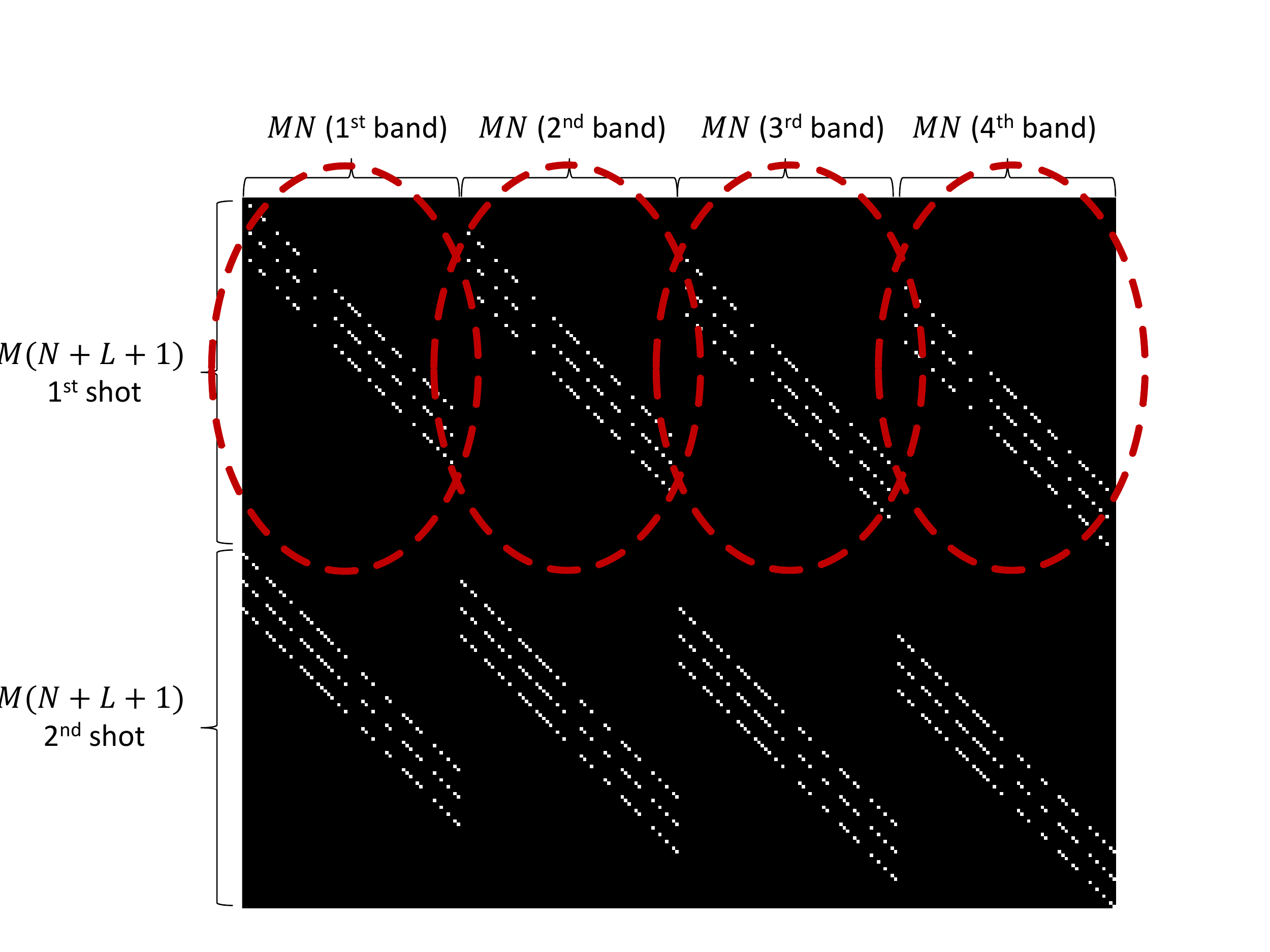}
\vspace*{0mm}
\caption{\sl 
The matrix ${\bf H}$ is presented for $M=N=8,L=4$, and $K=2$. The circled diagonal patterns that repeat horizontally correspond to the coded aperture pattern used in the first FPA shot. The second coded aperture pattern determines the next set of diagonals. Each FPA shot captures $M(N+L+1)=104$ measurements.
}
\vspace*{-5mm}
\label{fig.1}
\end{figure}

The coded aperture snapshot spectral imager (CASSI)~\cite{Wagadarikar2008single} is a compressive spectral imaging system 
that collects far fewer measurements than traditional spectrometers. In CASSI, ({\em i}) the 2D spatial information of a scene is coded by an aperture, ({\em ii}) the coded spatial projections are spectrally shifted by a dispersive element, and ({\em iii}) the coded and shifted projections are detected by a 2D focal plane array (FPA). For a 3D image cube of dimension $M\times N\times L$, the imaging process of CASSI can be written in a matrix-vector form,
\begin{equation}
\vspace*{0mm}
{\bf g} = {\bf H}{\bf f_0}+{\bf z},
\label{eq:CASSI}
\vspace*{0mm}
\end{equation}
where ${\bf f_0}\in\mathbb{R}^n$ is the vectorized 3D image cube of dimension $n=MNL$, and vectors ${\bf g}\in\mathbb{R}^m$ and ${\bf z}\in\mathbb{R}^m$ are the measurements and the additive noise, respectively.
The matrix ${\bf H}\in\mathbb{R}^{m\times n}$ models the linear relationship between ${\bf f_0}$ and ${\bf g}$,
and accounts for the effects of the coded aperture and the dispersive element.
Recently, Arguello et al.~\cite{Arguello2013higher} proposed a higher order model to characterize the CASSI system with greater precision. 
In this higher order CASSI model, each cubic voxel is shifted to an oblique voxel because of the continuous nature of the dispersion, and therefore the oblique voxel contributes to multiple measurements in the FPA. 
A sketch of the matrix ${\bf H}$ in the higher order CASSI model is depicted in Figure~\ref{fig.1}, where the image cube size is $M=N=8$, and $L=4$.
The matrix ${\bf H}$ consists of a set of $3$ diagonal patterns, accounting for the voxel energy impinging into 3 neighboring FPA pixels. 
The 3 diagonal patterns repeat in the horizontal direction, each time with a unit downward shift, as many times as the number of spectral bands. Each diagonal pattern is the coded aperture itself after being column-wise vectorized. Just below, the next set of diagonal patterns is determined by the coded aperture pattern used in the subsequent shot.
With $K$ shots of CASSI, the number of measurements is $m=KM(N+L+1)$ (see~\cite{Arguello2013higher} for details).

\vspace*{0mm}
\section{Proposed Algorithm}
\label{sec:Algo}
\vspace*{0mm}

The goal of our proposed algorithm is to reconstruct the image cube ${\bf f_0}$ from its compressive measurements ${\bf g}$, where the matrix ${\bf H}$ is known.
In this section, we describe our algorithm in detail. The algorithm employs ({\em i}) approximate message passing (AMP)~\cite{DMM2009}, an iterative algorithm for compressive sensing problems, and ({\em ii}) adaptive Wiener filtering, a hyperspectral image denoiser that can be applied within AMP.

{\bf Scalar channels:}
Below we describe that the linear imaging system model in~\eqref{eq:CASSI} can be converted to 3D image denoising in scalar channels. Therefore, we begin by defining scalar channels, where the noisy observations~${\bf q}$ of the image cube~${\bf f_0}$
obey 
${\bf q=f_0+v}$,
and ${\bf v}$ is the additive noise vector. Recovering ${\bf f_0}$ from ~${\bf q}$ is known as a 3D image denoising problem.

{\bf Approximate message passing:}
AMP~\cite{DMM2009} has recently become a popular algorithm for solving signal reconstruction problems in linear systems as
defined in~\eqref{eq:CASSI}. 
The AMP algorithm proceeds iteratively according to
\begin{align}
\vspace*{0mm}
{\bf f}^{t+1}&=\eta_t({\bf H}^T{\bf r}^t+{\bf f}^t)\label{eq:AMPiter1},\\
\vspace*{0mm}
{\bf r}^t&={\bf g}-{\bf Hf}^t+\frac{1}{R}{\bf r}^{t-1}
\langle\eta_{t-1}'({\bf H}^T{\bf r}^{t-1}+{\bf f}^{t-1})\rangle\label{eq:AMPiter2},
\vspace*{0mm}
\end{align}
where~${\bf H}^T$ is the transpose of ${\bf H}$, $R=m/n$ represents the measurement rate, $\eta_t(\cdot)$ is a denoising function at the $t$-th iteration, $\eta_t'({\bf s})=\frac{\partial}{\partial {\bf s}}\eta_t({\bf s})$, and~$\langle{\bf u}\rangle=\frac{1}{n}\sum_{i=1}^n u_i$
for some vector~${\bf u}=(u_1,u_2,\ldots,u_n)$.
The last term in~\eqref{eq:AMPiter2} is called the ``Onsager reaction term"~\cite{Thouless1977,DMM2009} in statistical physics.
In the~$t$-th iteration, we obtain the vectors~${\bf f}^t$ and~${\bf r}^t$. 
We highlight that the vector~${\bf H}^T{\bf r}^t+{\bf f}^t$ in~\eqref{eq:AMPiter1} can be regarded as a noise-corrupted version of~${\bf f_0}$ in the~$t$-th iteration with noise variance~$\sigma_t^2$, and therefore~$\eta_t(\cdot)$ is a 3D image denoising function that is performed on a scalar channel
${\bf q}^t = {\bf H}^T{\bf r}^t +{\bf f}^t= {\bf f_0} + {\bf v}^t$,
where the noise level $\sigma^2_t$ is estimated by~\cite{Montanari2012},
\begin{equation}
\vspace*{0mm}
\widehat{\sigma}^2_t=\frac{1}{m}\sum_{i=1}^m (r^t_i)^2,\nonumber
\vspace*{0mm}
\end{equation}
and~$r^t_i$ denotes the $i$-th component of the vector~${\bf r}^t$ in~\eqref{eq:AMPiter2}.

{\bf Adaptive Wiener filter:}
We are now ready to describe our 3D image denoiser, which is the function~$\eta_t(\cdot)$ in~\eqref{eq:AMPiter1}. 
First, we want to find a sparsifying transform such that hyperspectral images have only a few large coefficients in this transform domain, because based on the sparsifying coefficients, some shrinkage function can be applied in order to suppress noise~\cite{Donoho1994}.
Inspired by Arguello and Arce~\cite{Arguello2014},
we apply a wavelet transform to each of the 2D images in a 3D cube, and then apply a DCT along the spectral dimension. That is, the sparsifying transform ${\bf\Psi}$ can be expressed as a Kronecker product of a DCT transform ${\bf \Phi}$ and a 2D wavelet transform ${\bf W}$, i.e., ${\bf\Psi=\Phi\otimes W}$, and it can easily be shown that ${\bf\Psi}$ is an orthonormal transform. Our 3D image denoising is processed on the sparsifying coefficients~${\pmb{\theta}}_{{\bf q}}^t={\bf\Psi q}^t$.

Our proposed 3D image denoiser is a modification of the adaptive Wiener filter in our previous work~\cite{Tan_CompressiveImage2014}, which is inspired by M{\i}h{\c{c}}ak et al.~\cite{Kivanc1999}. 
Let ${\theta}_{{\bf q},i}^t$ denote the $i$-th element of~${\pmb{\theta}}_{\bf q}^t$. 
The coefficients $\widehat{\pmb{\theta}}_{\bf f}^t$ of the estimated (denoised) image cube ${\bf f}^t$ are obtained by Wiener filtering,
\begin{equation}
\vspace*{0mm}
\widehat{\theta}_{{\bf f},i}^{t+1}
=\frac{\max\{0,\widehat{\nu}_{i,t}^2-{\widehat{\sigma}_t^2}\}}{\widehat{\nu}_{i,t}^2}\left(\theta_{{\bf q},i}^t-\widehat{\mu}_{i,t}\right)+\widehat{\mu}_{i,t},
\label{eq:Wiener}
\vspace*{0mm}
\end{equation}
where $\widehat{\mu}_{i,t}$ and $\widehat{\nu}_{i,t}^2$ are the empirical mean and variance of ${\theta}_{{\bf q},i}^t$ within an appropriate wavelet subband, respectively. 
Note that in our previous work~\cite{Tan_CompressiveImage2014}, the variance was estimated locally from the coefficients within a $5\times 5$ window, and all coefficients had different variance values. In the current work, the variance is estimated from an entire subband, and the coefficients within each subband share the same variance. Such a denoiser has a simpler structure, and is likely to help prevent AMP from diverging.
Taking the maximum between 0 and $(\widehat{\nu}_{i,t}^2-{\widehat{\sigma}_t^2})$ ensures that if the empirical variance $\widehat{\nu}_{i,t}^2$ of the noisy coefficients is smaller than the noise variance ${\widehat{\sigma}_t^2}$, then the corresponding noisy coefficients are set to 0. After obtaining the denoised coefficients~$\widehat{\pmb{\theta}}_{\bf f}^{t+1}$ from ${\pmb{\theta}}_{{\bf q}}^t={\bf\Psi q}^t$, the estimated image cube in the $(t+1)$-th iteration satisfies ${\bf f}^{t+1}=\eta_t({\bf q}^t)={\bf \Psi}^{-1}\widehat{\pmb{\theta}}_{\bf f}^{t+1}={\bf\Psi}^T\widehat{\pmb{\theta}}_{\bf f}^{t+1}$.

{\bf AMP-3D-Wiener:}
It has been discussed~\cite{Tan_CompressiveImage2014} that when the sparsifying transform is orthonormal, the derivative calculated in the transform domain is equivalent to the derivative in the image domain.
According to~\eqref{eq:Wiener}, the derivative of the Wiener filter in the transform domain with respect to $\widehat{\theta}_{{\bf q},i}^t$ is~$\max\{0,\widehat{\nu}_{i,t}^2-\widehat{\sigma}_t^2\}/\widehat{\nu}_{i,t}^2$. Because the sparsifying transform ${\bf\Psi}$ is orthonormal, the Onsager term~\eqref{eq:AMPiter2} can be calculated as

\begin{equation}
\vspace*{0mm}
\langle\eta'_t({\bf H}^T{\bf r}^t + {\bf f}^t)\rangle = \frac{1}{n} \sum_i\frac{\max\{0,\widehat{\nu}_{i,t}^2-\widehat{\sigma}_t^2\}}{\widehat{\nu}_{i,t}^2}.\nonumber
\vspace*{0mm}
\end{equation}

Basic AMP has been proved to converge with i.i.d. Gaussian matrices 
and scalar functions $\eta_t(\cdot)$~\cite{Bayati2011}, i.e., $\widehat{\theta}^t_{{\bf f},i}$ only depends on its corresponding noisy coefficient $\widehat{\theta}^t_{{\bf q},i}$. Other AMP variants~\cite{Swamp2014,Rangan2014ISIT,RanganADMMGAMP2015} have been proposed in order to encourage convergence for a broader class of measurement matrices.
The matrix ${\bf H}$ defined in~\eqref{eq:CASSI} is not i.i.d. Gaussian, but highly structured as shown in Figure~\ref{fig.1},
and the adaptive Wiener filter in~\eqref{eq:Wiener} is not a scalar function owing to $\widehat{\mu}_{i,t}$ and $\widehat{\nu}_{i,t}^2$ being functions of multiple noisy coefficients~$\widehat{\theta}^t_{{\bf q},i}$.
Unfortunately, 
AMP-3D-Wiener encounters divergence issues with this matrix~${\bf H}$.
We choose to apply ``damping"~\cite{Rangan2014ISIT,Vila2014},
which resembles a technique
used in Gaussian belief propagation~\cite{Johnson2009}, 
to solve for the divergence problems of AMP-3D-Wiener, because it is simple and only increases the runtime modestly. Specifically, damping is an extra step within AMP iterations that updates the values of~${\bf r}^t$ and~${\bf f}^{t+1}$ by weighted sums as shown in Lines $2$ and $7$ of Algorithm~\ref{algo:amp_wiener}.
We will show in Section~\ref{sec:NumSim} that AMP-3D-Wiener converges with an appropriate amount of damping, and AMP-3D-Wiener serves as a demonstration that non-scalar denoisers have promises in the AMP framework~\cite{Tan_CompressiveImage2014,Donoho2013}.

\vspace*{0mm}
\begin{algorithm}[h]
\caption{AMP-3D-Wiener}
\label{algo:amp_wiener}
\textbf{Inputs:} ${\bf g}$, ${\bf H}$, $0<\alpha\le1$, maxIter\\
{\bf Outputs:} $\widehat{{\bf f}}_\text{AMP}$\\
\textbf{Initialization:} ${\bf f}^1={\bf 0}$, ${\bf r}^{0}={\bf 0}$
\begin{algorithmic}
\For{$t=1:\text{maxIter}$}\\
\begin{enumerate}
\item
${\bf r}^t={\bf g}-{\bf Hf}^t+\frac{1}{R}{\bf r}^{t-1}
\frac{1}{n} \sum_{i=1}^n \frac{\max\{0,\widehat{\nu}_{i,t-1}^2-\widehat{\sigma}_{t-1}^2\}}{\widehat{\nu}_{i,t-1}^2}$
\item
${\bf r}^t = \alpha\cdot{\bf r}^t + (1-\alpha)\cdot{\bf r}^{t-1}$
\item
${{\bf q}}^t={\bf H}^T{\bf r}^t+{\bf f}^t$
\item
$\widehat{\sigma}^2_t=\frac{1}{m}\sum_j ({r}^t_j)^2$
\item
${\pmb{\theta}}_{\bf q}^t = {\bf\Psi}{\bf q}^t$
\item
$\widehat{\theta}_{{\bf f},i}^t=\frac{\max\{0,\widehat{\nu}_{i,t}^2-{\widehat{\sigma}_t^2}\}}{\widehat{\nu}_{i,t}^2}\left(\theta_{{\bf q},i}^t-\widehat{\mu}_{i,t}\right)+\widehat{\mu}_{i,t}$
\item
${\bf f}^{t+1}=\alpha\cdot{\bf\Psi}^T\widehat{\pmb{\theta}}_{\bf f}^t+(1-\alpha) \cdot{\bf f}^{t}$
\end{enumerate}
\EndFor\\
$\widehat{{\bf f}}_\text{AMP}={\bf f}^{\text{maxIter+1}}$
\end{algorithmic}
\vspace*{0mm}
\end{algorithm}

\vspace*{0mm}
\section{Numerical Results}
\label{sec:NumSim}
\vspace*{0mm}

In this section, we compare the reconstruction quality and runtime of AMP-3D-Wiener,
gradient projection for sparse reconstruction (GPSR)~\cite{GPSR2007}, and two-step iterative shrinkage/thresholding (TwIST)~\cite{Wagadarikar2008,NewTWIST2007}.
In all experiments, we use the same coded aperture pattern for AMP-3D-Wiener, GPSR, and TwIST.
In order to quantify the reconstruction quality of each algorithm, the peak signal to noise ratio (PSNR) of each 2D slice in reconstructed cubes is measured.

In AMP, the damping parameter~$\alpha$ is set to be 0.2. The choice of damping mainly depends on the structure of the imaging model in~\eqref{eq:CASSI} but not on the characteristics of the image cubes, and thus the value of the damping parameter~$\alpha$ need not be tuned in our experiments.

To reconstruct the image cube~${\bf f_0}$, GPSR and TwIST minimize objective functions of the form
$
\widehat{\bf f}=\arg\min_{\bf f} \frac{1}{2}\|{\bf g-Hf}\|_2^2 + \beta\cdot\phi({\bf f})
$,
where $\phi(\cdot)$ is a regularization function that characterizes the structure of the image cube ${\bf f_0}$, and $\beta$ is a regularization parameter that balances the weights of the two terms in the objective function. In GPSR, $\phi({\bf f}) = \|{\bf \Psi f}\|_1$; in TwIST, $\phi({\bf f})$ is the total variation function~\cite{Chambolle2004}.
We select the optimal values of $\beta$ for GPSR and TwIST manually, i.e., we run GPSR and TwIST with $5-10$ 
different values of $\beta$, and select the results with the highest PSNR.


In order to compare the performance of AMP-3D-Wiener, GPSR, and TwIST, an image cube is experimentally acquired using a wide-band Xenon lamp as the illumination source, modulated by a visible monochromator spanning the spectral range between $448$ nm and $664$ nm, and each waveband has $9$ nm width. The image intensity was captured using a grayscale CCD camera, with pixel size $9.9$ $\mu$m, and 8 bits of intensity levels. The resulting test image cube is of size $M \times N = 256 \times 256$, and $L=24$.

{\bf Setting 1:} The measurements~${\bf g}$ are captured with $K=2$ shots such that the two coded apertures are complementary. Therefore, we ensure that the norm of each column in ${\bf H}$~\eqref{eq:CASSI} is similar. 
The measurement rate is $m/n=KM(N+L+1)/(MNL)\approx0.09$.
Moreover, we add zero-mean Gaussian noise ${\bf z}$ to the measurements such that the signal to noise ratio (SNR) is 20 dB. The SNR is defined as $10\log_{10}(\mu_g/\sigma_\text{noise})$~\cite{Arguello2014}, where $\mu_g$ is the mean value of the measurements ${\bf Hf_0}$ and $\sigma_\text{noise}$ is the standard deviation of~${\bf z}$.

\begin{figure}[t]
\vspace*{+3mm}
\begin{center}
\includegraphics[width=70mm]{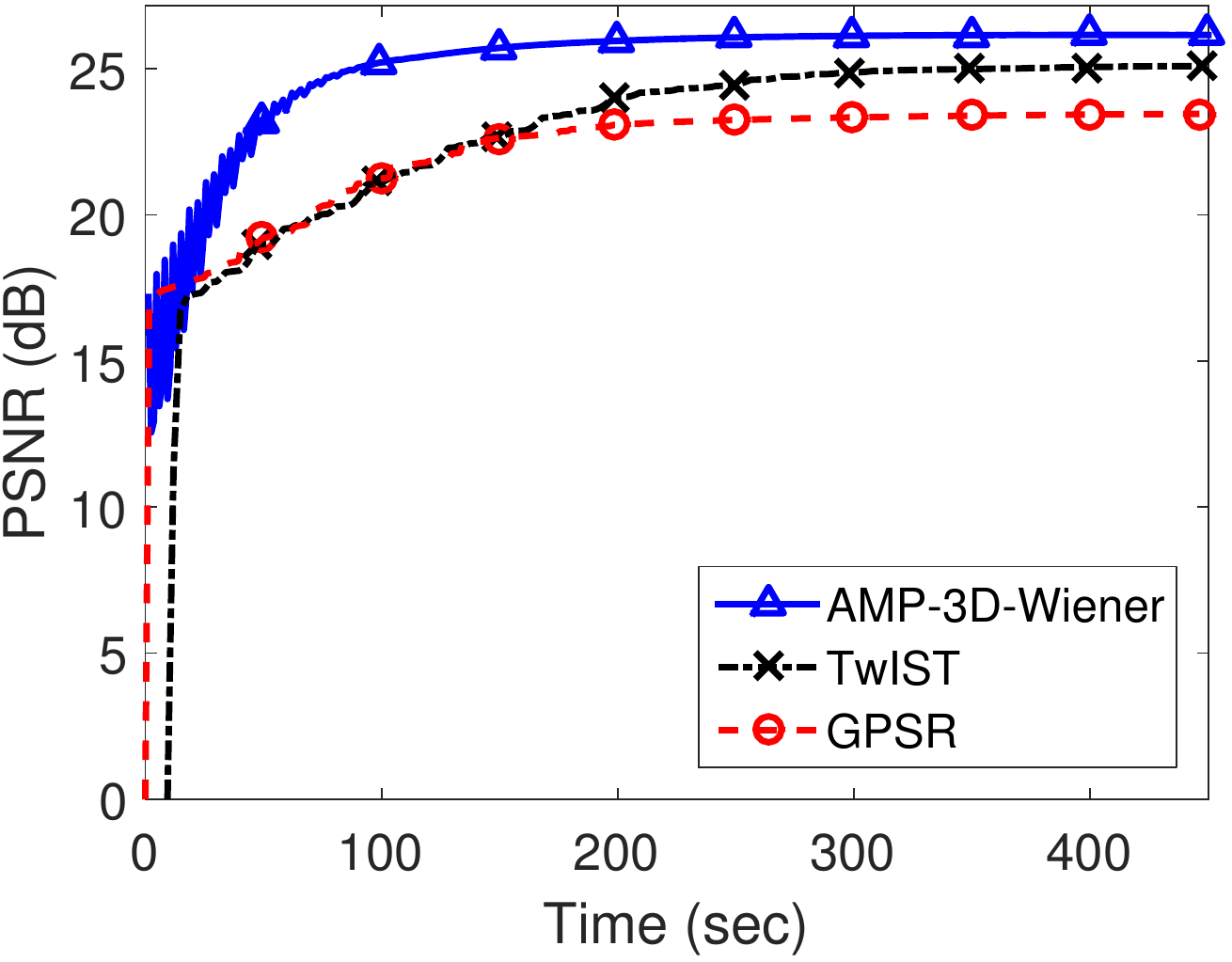}
\end{center}
\vspace*{0mm}
\caption{\sl 
Runtime versus average PSNR comparison of AMP-3D-Wiener, GPSR, and TwIST. Cube size is $M=N=256$, and $L=24$. The measurements are captured with $K=2$ shots using complementary coded apertures, and the number of measurements is $m=143,872$.}
\vspace*{0mm}
\label{fig.iter_Psnr}
\end{figure}

\begin{figure}[h]
\vspace*{-3mm}
\begin{center}
\includegraphics[width=85mm]{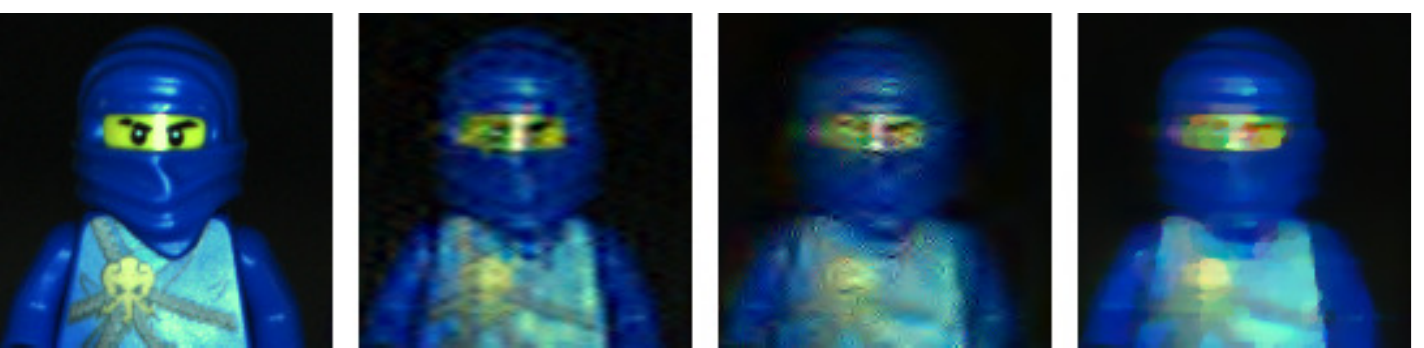}
\end{center}
\vspace*{0mm}
\caption{\sl 
Moving from left to right, the panels correspond to the groundtruth and image cubes reconstructed by AMP-3D-Wiener, GPSR, and TwIST. 
The 2D RGB images shown in this figure are converted from their corresponding 3D image cubes.
(The target object presented in the experimental results was not endorsed by the trademark owners and it is used here
as fair use to illustrate the quality of reconstruction of compressive spectral image measurements. LEGO is a trademark of the LEGO Group,
which does not sponsor, authorize or endorse the images in this paper. The LEGO Group. All Rights Reserved. http://aboutus.lego.com/enus/
legal-notice/fair-play/.)}
\vspace*{-7mm}
\label{fig.All_10}
\end{figure}

Figure~\ref{fig.iter_Psnr} compares the reconstruction quality of 
AMP-3D-Wiener, 
GPSR, 
and TwIST
within $450$ seconds. 
Runtime is measured on a Dell OPTIPLEX 9010 running an Intel(R)
CoreTM i7-860 with 16GB RAM, and the environment is Matlab R2013a.
In Figure~\ref{fig.iter_Psnr},
the horizontal axis represents runtime in seconds, and the vertical axis is the averaged PSNR over the 24 spectral bands.
Although the PSNR of AMP-3D-Wiener oscillates during the first few iterations, which may be because the matrix~${\bf H}$ is ill-conditioned, it becomes stable after 50 seconds and reaches a higher level compared to the PSNRs of GPSR and TwIST at 50 seconds.
After 450 seconds, the average PSNRs of the cubes reconstructed by AMP-3D-Wiener, GPSR, and TwIST are $26.16$ dB, $23.46$ dB and $25.10$ dB, respectively. Figure~\ref{fig.All_10} displays the reconstructed cubes in the form of 2D RGB images, and we can see that AMP-3D-Wiener produces images with better quality; images reconstructed by GPSR and TwIST are blurrier.

Note that in 450 seconds, AMP-3D-Wiener and GPSR run roughly $400$ iterations, while TwIST runs around $200$ iterations. Therefore, for the rest of the simulations, we run AMP-3D-Wiener and GPSR with 400 iterations, and TwIST with 200 iterations, so that all algorithms complete within the similar amount of time.

\begin{figure}[t]
\vspace*{+3mm}
\begin{center}
\includegraphics[width=70mm]{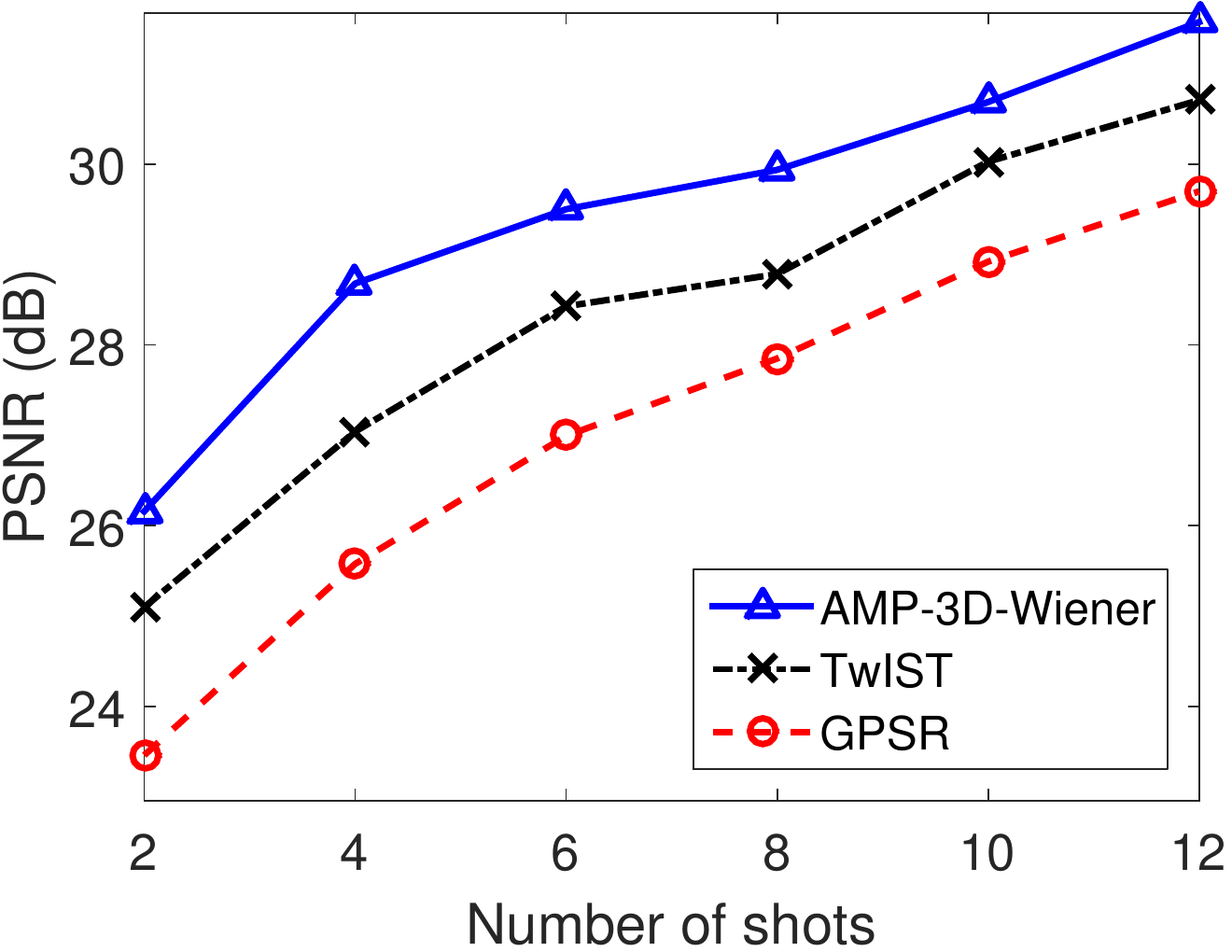}
\end{center}
\vspace*{0mm}
\caption{\sl 
Number of shots versus average PSNR comparison of AMP-3D-Wiener, GPSR, and TwIST. Cube size is $N=M=256$, and $L=24$. The measurements are captured using pairwise complementary coded apertures.
}
\vspace*{-7mm}
\label{fig.shots_Psnr}
\end{figure}

{\bf Setting 2:} In Setting 1, the measurements are captured with $K=2$ shots. We now test our algorithm on the setting where the number of shots varies from $K=2$ to $K=12$ with pairwise complementary coded apertures. 
Specifically, we randomly generate the coded aperture for the $k$-th shot for $k=1,3,5,7,9,11$, and the coded aperture in the $(k+1)$-th shot is the complement of the aperture in the $k$-th shot.
In this setting, a moderate amount of noise (20 dB) is added to the measurements. Figure~\ref{fig.shots_Psnr} presents the PSNR changes of the reconstructed cubes as the number of shots increases, and AMP-3D-Wiener consistently beats GPSR and TwIST.

{\bf Test on natural scenes:}
We have also tested our algorithm on the dataset ``natural scenes 2004" ~\cite{Foster2006,Foster2006url}.
In this dataset, there are 8 image cubes with $L=33$ spectral bands and spatial resolution of around $1000\times 1000$.
To satisfy the dyadic constraint of the 2D wavelet, we crop their spatial resolution to be $M=N=512$.
The measurements are captured with $K=2$ shots, and the measurement rate is $m/n=KM(N+L+1)/(MNL)\approx0.065$.
We test for measurement noise levels such that the SNRs are 15 dB and 20 dB. 
The typical runtimes for AMP with 400 iterations, GPSR with 400 iterations, and TwIST with 200 iterations are approximately $2,800$ seconds.
We run the algorithms on $5$ different complementary coded apertures, and
the average PSNR for each algorithm is shown in Table~\ref{tb:scene2004}. We highlight the highest PSNR among AMP-3D-Wiener, GPSR, and TwIST using bold fonts.  It can be seen from Table~\ref{tb:scene2004} that AMP-3D-Wiener usually outperforms GPSR by $2-5$ dB in terms of the PSNR, and outperforms TwIST by $0.2- 4$ dB.

\begin{table}[h]
\vspace*{0mm}
\centering
\begin{tabular}{|c|| c | c | c ||c|c|c|}
\hline
&\multicolumn{3}{|c||}{15 dB}&\multicolumn{3}{|c|}{20 dB}\\
\hline
& { AMP} & GPSR & TwIST & { AMP} & GPSR & TwIST \\
\hline 
Scene 1& {\bf30.46} & 28.43 &30.14 & {\bf30.54} & 28.52 & 30.27 \\

Scene 2& {\bf27.33} & 24.74 &27.09 & {\bf27.74} &  24.88 &  27.42\\

Scene 3& {\bf33.29} & 29.53 & 31.87 & {\bf33.10} & 29.57 & 31.93\\
Scene 4& {\bf32.04} & 27.00 & 31.61 & {\bf32.25} & 27.22 & 31.97\\
Scene 5& {\bf27.44} & 24.28 & 26.46 & {\bf27.80} & 24.64 & 26.84\\
Scene 6& {\bf29.17} & 25.02 & 25.82 & {\bf30.06} & 25.55 & 26.27 \\
Scene 7& {\bf36.36} & 33.07 & 33.76 & {\bf37.14} & 33.54 & 34.21\\
Scene 8& {\bf32.15} & 28.19& 28.17 & {\bf32.99} & 28.79 & 28.57 \\
\hline
\end{tabular}     
\caption{\sl 
Average PSNRs of AMP-3D-Wiener, GPSR, and TwIST for the dataset ``natural scene 2004"~\cite{Foster2006}. The spatial dimensions of the cubes are cropped to $M=N=512$, and each cube has $L=33$ spectral bands. 
The measurements are captured with $K=2$ shots, and the number of measurements is $m=559,104$.
}
\vspace*{0mm}
\label{tb:scene2004}
\end{table}

\vspace*{0mm}
\section{Conclusion}
\label{sec:con}
\vspace*{0mm}

In this paper, we considered compressive hyperspectral imaging reconstruction in coded aperture snapshot spectral imager (CASSI) systems.
Considering that the CASSI system is a great improvement in terms of imaging quality and acquisition speed over conventional spectral imaging techniques, it is desirable to further improve CASSI by accelerating the 3D image cube reconstruction process. Our proposed AMP-3D-Wiener used an adaptive Wiener filter as a 3D image denoiser within the approximate message passing (AMP)~\cite{DMM2009} framework. AMP-3D-Wiener was faster than existing image cube reconstruction algorithms, and also achieved better reconstruction quality.

\vspace*{0mm}
\section*{Acknowledgments}
\vspace*{0mm}
We thank S. Rangan and P. Schniter for inspiring discussions on approximate message passing; L. Carin, and X. Yuan for kind help on numerical experiments; J. Zhu for informative explanations about CASSI systems; and N. Krishnan for detailed suggestions on the manuscript.

\ifCLASSOPTIONcaptionsoff
\newpage
\fi
\bibliographystyle{IEEEtran}
\bibliography{IEEEabrv,cites}

\end{document}